%% file: report.tex
\documentclass[twocolumn,tighten,twocolappendix]{aastex6}

\usepackage{amsmath}
\usepackage{xspace}

\newcommand{\br}[1]{\ensuremath{\left[ #1 \right]} }
\newcommand{\rbr}[1]{\ensuremath{\left( #1 \right)} }
\newcommand{\sbr}[1]{\ensuremath{\left\{ #1 \right\}} }

\newcommand{\avg}[1]{\ensuremath{\left< #1 \right>} }
\newcommand{\E}[1]{\ensuremath{\times 10^{#1}} }
\newcommand{\Ex}[1]{\ensuremath{\mathbb E\br{#1}}}
\newcommand{\Var}[1]{\ensuremath{\mathbb V\br{#1}}}

\newcommand{\msol}{\ensuremath{M_{\odot}}\xspace}

\newcommand{\maxi}{MAXI J0911\xspace}
\newcommand{\xmm}{\textit{XMM-Newton}\xspace}
\newcommand{\nustar}{\textit{NuSTAR}\xspace}
\newcommand{\low}{\ensuremath{\ell}ow\xspace}
\newcommand{\LFT}{\ensuremath{\mbox{LF}_2}\xspace}

\begin{document}

\title{%
The stochastic X-ray variability of the accreting millisecond pulsar MAXI J0911--655
}

\AuthorCallLimit=1

\author{Peter Bult}

\affil{
Astrophysics Science Division,
NASA Goddard Space Flight Center,
Greenbelt, MD 20771
}

\begin{abstract} 
    In this work I report on the stochastic X-ray variability of the 340 Hz accreting 
    millisecond pulsar MAXI J0911--655. 
    Analyzing pointed observations of the \textit{XMM-Newton} and \textit{NuSTAR} 
    observatories I find that the source shows broad band-limited stochastic variability 
    in the $0.01-10$ Hz range, with a total fractional variability of $\sim24\%$ rms in
    the $0.4-3$ keV energy band, which increases to $\sim40\%$ rms in the $3-10$ keV
    band. Additionally a pair of harmonically related quasi-periodic oscillations 
    are discovered. The fundamental frequency of this harmonic pair is observed 
    between frequencies of $62$ mHz and $146$ mHz. Like the band-limited noise,
    the amplitude of the QPOs show a steep increase as a function of energy,
    suggesting they share a similar origin, which is likely the inner accretion flow.
    Based on their energy dependence and their frequency relation with respect to 
    the noise terms, the QPOs are identified as Low-Frequency oscillations, and 
    discussed in terms of Lense-Thirring precession model.
\end{abstract}


\section{Introduction} 

Accreting Millisecond X-ray Pulsars (AMXPs) are a class of transient neutron
star Low-Mass X-ray Binaries (LMXBs) that show coherent pulsations during their
outbursts (see \citealt{Patruno2012b} for a review). Such pulsations are
attributed to magnetically channeled accretion onto the neutron star, so that
emission from a localized impact region gives rise to periodic intensity
variations at the neutron star spin frequency.  By tracking the arrival time of
the pulsations the neutron spin frequency and its evolution can be measured.
This then gives a direct tool through which the torque mechanisms acting
between the star and the surrounding accretion flow may be studied
\citep{Psaltis1999b, Bildsten1998, Haskell2011}.  Additionally, through the
timing of the pulsar the binary orbit and evolution may be investigated
\citep{Patruno2012a}, while careful modeling of the pulse waveform may be used
to extract information on otherwise elusive neutron star properties, such as
mass, radius, and magnetic field strength \citep{Poutanen2003, Leahy2008,
Psaltis2014}.

In addition to coherent pulsations, the X-ray emission from AMXPs also
shows rich stochastic variability. Like the broader class of LMXBs
\citep{Klis2006}, various timing features may be distinguished in AMXP
light curves, including broad band-limited noise terms, and narrow
Quasi-Periodic Oscillations (QPOs). Furthermore, the morphology,
relative frequencies, and correlations with luminosity or energy
spectra that may be observed for these timing features are all largely 
consistent with those observed in the atoll class of accreting neutron stars
\citep{Straaten2005}. 

Atoll type neutron stars are named for the pattern they trace out
in the color-color diagram as their luminosity changes
\citep{Hasinger1989}. At high luminosity, their energy spectrum is soft and
the bulk of their variability features narrow and concentrated at high 
frequencies ($>10$ Hz). As the luminosity varies, atoll sources trace out a banana
shaped pattern in the color-color diagram, which is usually
sub-categorized into three source states (\textit{upper-, lower-} and
\textit{lower-left banana}) depending on the specific morphology
of the power spectrum. For lower luminosities, atolls transition into
the \textit{island state}, which is characterized by a harder
energy spectrum. Meanwhile, the power spectral features shift to lower
frequencies ($1-100$ Hz), while gaining in both width and amplitude.
This trend continues to the lowest observed luminosities where
such sources may enter an \textit{extreme island state}. Here
the energy spectrum is dominated by a hard power law, while the
bulk power density has shifted down to $0.1-10$ Hz with only
weak QPOs or none at all.

Given that the stochastic timing signatures are generally attributed 
to the accretion flow, it is of interest to compare the differences 
and similarity between pulsating and non-pulsating objects as this
offers a path to investigating the coupling mechanisms between the 
neutron star and the accretion flow.
In this work I therefore report on the first stochastic X-ray 
variability study of the accreting millisecond pulsar MAXI J0911--655
(henceforth \maxi) based on observed of the \xmm and \nustar
observatories.

\subsection{MAXI J0911--655}
The X-ray transient \maxi, was discovered on
February 19th, 2016 \citep{Serino2016} with the \textit{MAXI/GSC}. The source
was immediately associated with the globular cluster NGC 2808, a position
that was later confirmed by \textit{Swift/XRT} \citep{Kennea2016} and \textit{Chandra}
\citep{Homan2016} observations. Subsequent monitoring with \textit{INTEGRAL} and
\textit{Swift/XRT} has shown that \maxi has remained active, showing a persistent
flux of about $7$ mCrab \citep{Ducci2016}. At the time of writing this source is yet to 
transition into quiescence, placing the duration of the outburst at
approximately one year.

The nature of the compact object in \maxi was settled when its 
340 Hz pulsations were 
discovered by \citet{Sanna2017a}. Using pointed \xmm and \nustar
observations, these authors studied the pulsations and showed the AMXP
is set in a compact binary with a 44.3 minutes orbital period and a 
companion star of $>0.024$ \msol. The nature of the companion is not
definitively constrained, but is likely either a hot, helium white dwarf,
or an old brown dwarf. Additionally, they found that the energy spectrum 
of \maxi is relatively hard, with a $\Gamma=1.7$ power-law dominating over
a weak $kT\sim0.5$ blackbody component.

\section{Data Reduction}
\label{sec:datareduction}

\subsection{XMM-Newton}
    In this work I analyze the \textsc{epic-pn} data of two \xmm observations
    of \maxi.  The first observation took place on April 24, 2016 (ObsID
    0790181401) and the second on May 22, 2016 (ObsID 0790181501). For both
    observations the \textsc{epic-pn} camera was operated in \textsc{timing}
    mode, yielding event list data at a time resolution of 29.56 $\mu s$.

    To obtain science grade products, the data was processed with \textsc{sas} 
    version 15.0.0 using the most recent calibration files available.
    Standard data screening criteria were applied to the data, selecting
    only those events in the $0.4-10$ keV energy range with 
    \textsc{pattern}$\leq4$ and screening \textsc{flag}$=0$.
    The source data was extracted from a 15 bin wide rectangular region 
    with \textsc{rawx} coordinates $[31:45]$.
    Finally the \textsc{sas} tool \textsc{barycen} was applied to correct 
    the event arrival times to the Solar System barycenter, using the source 
    coordinates of \citet{Homan2016}.

    The background estimate was obtained similarly, but from a 3 bin wide region
    with the \textsc{rawx} coordinates $[3:5]$. Since this region is smaller 
    than the source extraction region I multiplied the resulting background
    count rate by a factor of 5 to ensure both source and background rates 
    reflect a comparable effective area.

    The source count rates obtained are 40(1) and 35(2) counts/s for the first
    and second observations, respectively. The associated background count
    rates are respectively 0.8(4) and 0.4(2) counts/s.
    
\subsection{NuSTAR}
    \nustar observed \maxi on May 24, 2016 (ObsID 90201024002),
    and again 180 days later on November 23, 2016 (ObsID 90201042002).
    The data was processed using the \textsc{nustardas} software pipeline
    version 1.6.0. This was done separately for each of the two focal plane
    modules (FPMA and FPMB).
    
    The source events in the $3-79$ keV energy band were extracted from a
    circular region with a $50"$ radius that was centered on the image source
    position. The filtered event arrival times were then corrected to the Solar
    System barycenter using the \textsc{barycorr} tool, again based on the
    source coordinates of \citet{Homan2016}. The background events were
    extracted similarly, but with a source region centered in the background
    field.  The resulting source count rates 
    are 2.0(2) counts/s and 3.0(2) counts/s, for the first and second observations, 
    respectively, with a background count rate of 0.01(1) counts/s in both
    cases.

\subsection{Timing Analysis} 
\label{sec:timing.analysis} 
    A stochastic timing analysis is performed on the cleaned and processed
    X-ray data. First all the arrival times are adjusted for the binary motion
    of neutron star based on the ephemeris of \citet{Sanna2017a}.  The event
    lists are then binned into $\sim250$ second long light curves using a time
    resolution of 480 $\mu s$ (the length of the time series is adjusted so
    that the number of time bins is a power of two).  For each of these light
    curves I compute the Fourier transform, from which a Leahy-normalized power
    spectrum is calculated \citep{Leahy1983b}.  These power spectra have a
    frequency resolution of $\sim0.004$ Hz and a limiting Nyquist frequency of
    $1040$ Hz.  All individual power spectral estimates are averaged per ObsID
    and corrected for Poisson noise.
    
    For the \xmm data the Poisson noise spectrum is modeled as a constant. That
    noise power is measured between $400$ Hz and $1000$ Hz, where no signal
    features are observed, and then subtracted from the averaged power density
    spectrum.

    In the case of the \nustar data the instrument deadtime needs to be
    taken into account, as it has a comparatively long timescale of
    $\sim2.5$ ms \citep{Harrison2013}. An elegant way of dealing with the \nustar deadtime
    was proposed by \citet{Bachetti2015} and involves computing the
    complex-valued cross spectrum from the two FPM light curves. The
    real component of the cross spectrum, also known as the
    co-spectrum, can then be adopted as an estimator of the source
    power density spectrum. 
    Because, to good approximation\footnote{%
        Deadtime is a multiplicative noise term, so the modulation imposed
        on the power spectrum is, to some degree, correlated with the
        spectrum of the signal itself.
        Correlating a multi-detector system can therefore not 
        completely remove deadtime effects. 
    },
    the deadtime is independent between detectors, while the signal itself
    is wholly correlated, this approach effectively filters out any deadtime modulation
    without the need to model it. 
    
    I would point out, however, that the gain achieved through considering
    the co-spectrum does not come for free. The cost of correlating the two
    FPM light curves is that the resulting signal-to-noise is lower than what 
    would have been obtained if the two light curves were simply added in the 
    time domain. Specifically, the uncertainties on the co-spectral powers are 
    larger than those obtained from a regular power spectrum by up to a factor of 
    $\sqrt{2}$  (see appendix \ref{app:nustar} for details).  
    
    Given that the observed event rate (0.5 s/count) is much larger than the 
    timescale of the deadtime, the effect of the dead time will be relatively 
    small, and the power spectrum will only be sensitive at comparatively low
    frequencies. I would therefore argue that in this case it is more appropriate 
    to add the two FPM light curves for the increased sensitivity, rather than 
    correlating them to reduce the deadtime modulation. 
    
\begin{figure}
        \centering
        \includegraphics[width=\linewidth]{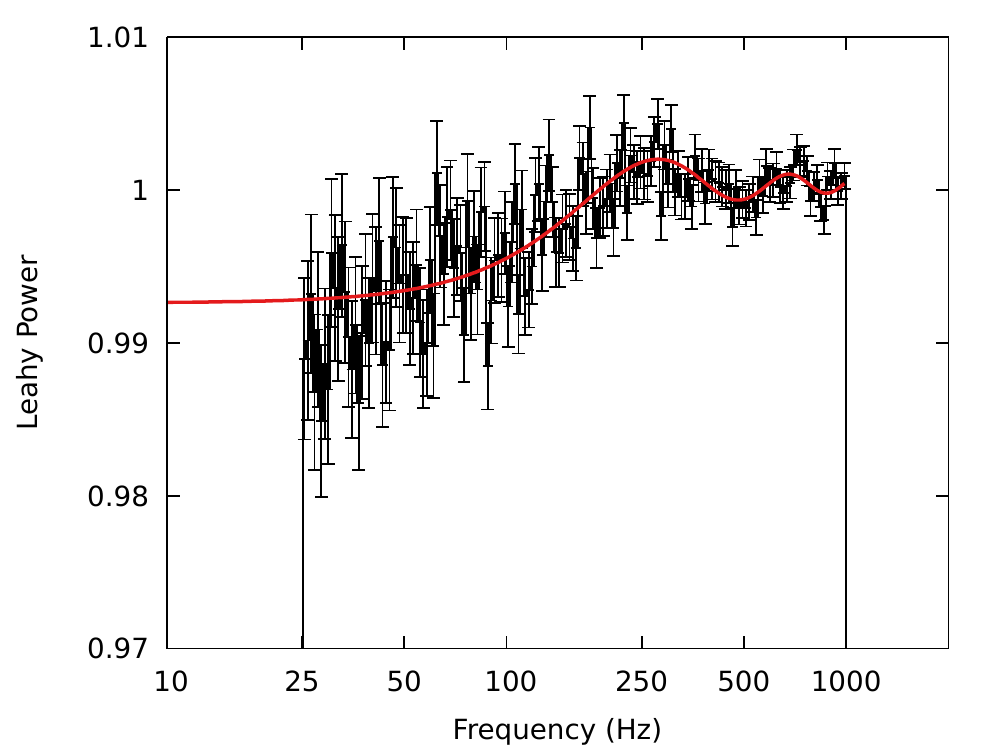}
        \caption{Deadtime model fitted to the \nustar data. See section
        \ref{sec:timing.analysis} for details.}
        \label{fig:deadtime}
    \end{figure}

    The \nustar power spectra are constructed by adding the concurrent
    FPM data. For each light curve the Fourier transforms are computed and
    normalized, and subsequently averaged in the complex domain.
    From this averaged Fourier transform a Leahy normalized power spectrum
    constructed. Note that this power spectrum will have an expected Poisson 
    noise level of $1$ (see appendix \ref{app:nustar}). All segments of the 
    ObsID are averaged to a single power spectrum. This power spectrum
    shows some deadtime modulation, which is modeled heuristically by taking 
    all powers above $25$ Hz and fitting the scaled sinc function 
    \[
        f\rbr{\nu \mid A, B, t_D} = A - 2 B t_D \mbox{sinc}\rbr{2 \pi \nu t_D},
    \]
    where $A$ and $B$ are a mean and scale parameter, respectively, and
    $t_D$ is the characteristic deadtime. 
    For the first \nustar observation
    the best fit gives
    $\chi^2/dof = 147/158$ with parameters $A=1\pm1\E{-4}$, $B=1.5\pm0.1$ and
    $t_D = 2.56 \pm 0.06$ ms (Figure \ref{fig:deadtime}). Below 10 Hz this curve
    is near constant at a level of $P_{\rm noise} \simeq A - 2B t_D = 0.992$, which 
    I adopt as the Poisson noise level and subtract from the data.
    The second \nustar observation has similar results, and differs only in
    the scale, $B=2.32\pm0.14$, so that the Poisson noise level is marginally
    higher, $P_{\rm noise} \simeq 0.994$.

    Finally, the Poisson noise corrected averaged power spectra are 
    renormalized to squared fractional rms, 
    and subsequently described in terms of a multi-Lorentzian 
    model \citep{Belloni2002}. Each component, $L\rbr{ \nu \mid r, Q, \nu_{\rm max} }$, is a 
    function of Fourier frequency $\nu$ and characterized by three parameters.
    Here $\nu_{\rm max} = \nu_0 \sqrt{ 1 + 1/4Q^2}$ is the characteristic frequency, 
    and $Q = \nu_0 / W$ the quality factor, where $\nu_0$ gives the centroid 
    frequency and $W$ is the full-width-at-half-maximum. The amplitude of
    an individual component is expressed in terms of fractional rms, $r$, 
    defined as
    \[
        r^2 = P = \int_0^{\infty} L(\nu) d\nu,
    \]
    where $P$ is the integrated power. 
    A component is considered to be significant if its integrated power has
    a single trial significance greater than three, that is, if $P/\sigma_P 
    \geq 3$.

\section{Results} 
\label{sec:results}

Because the \xmm and \nustar energy bands overlap, the \xmm data is split
in a low energy ($0.4-3$ keV) and high energy ($3-10$ keV) band, so that
the latter is covered by both observatories, allowing for a direct comparison
of their power spectra.

\begin{figure}[t]
    \centering
    \includegraphics[width=\linewidth]{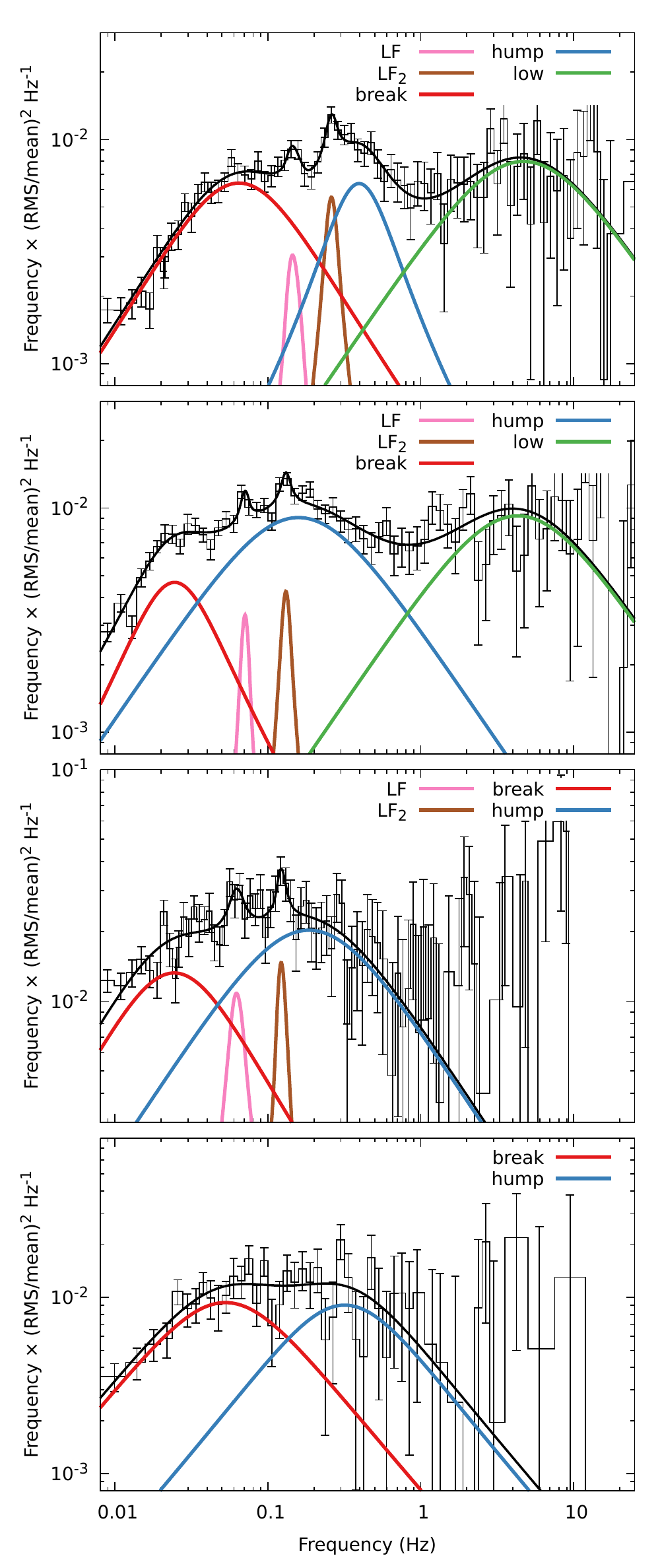}
    \caption{%
        Power density spectra for the XMM-1 (top) 
        and XMM-2 (second) data in the $0.4-3$ keV energy band,
        the NuSTAR-1 data (third), and the NuSTAR-2 data (bottom)
        in the $3-10$ keV energy band.
        Fit parameters are given in Table \ref{tab:fit}.
    }
    \label{fig:pds12}
\end{figure}

The first \xmm observation (XMM-1) shows significant power in the
$0.01$ to $10$ Hz frequency range (Figure \ref{fig:pds12}, top panel).
The integrated fractional rms amplitude is $24\%$ in the $0.4-3$ keV
band, and increases to $43\%$ in the $3-10$ keV band.

\begin{table*}
    \newcommand{\mc}[1]{\multicolumn2c{#1}}
    \centering
	\caption{%
    	Power spectrum fit parameters
		\label{tab:fit}	
	}
	\begin{tabular}{ l l D D D c }
    \decimals
	\tableline
    Energy& Component & \mc{Frequency} &\mc{Quality}& \mc{Amplitude} & $\chi^2$ / dof \\
    (keV) &           & \mc{(Hz)}      &\mc{~}      & \mc{(\% rms)}  &          \\
    \cutinhead{XMM-1} 
    ~       &  break & 0.065(6)  &  0.18(6)    &  13.1(6)  &   \\
    ~       &  hump  & 0.40(5)   &  0.8(3)     &  9.8(1.8) &   \\
    $0.4-3$ &  \low  & 4.7(9)    &  0. ~(fixed)&  15.8(8)  & 70 / 81   \\
    ~       &  LF    & 0.146(5)  &  4.2(1.8)   &  3.3(5)   &   \\
    ~       &  \LFT  & 0.260(7)  &  4.3(1.1)   &  4.4(4)   &   \\
    \tableline
    ~       &  break & 0.046(6)  &  0.35(11)   &  15.3(1.8) & \\
    $3-10$  &  hump  & 0.29(2)   &  0.26(15)   &  23.4(1.9) & 104/87 \\ 
    ~       &  \low  & 6.4(1.6)  &  0. ~(fixed)&  28.5(1.9) & \\
    \cutinhead{XMM-2} 
    ~       & break & 0.0245(18) &  0.52(9)     &  9.4(9)  &  ~  \\
    ~       & hump  & 0.16(2)    &  0. ~(fixed) & 16.9(7)  &  ~  \\
    $0.4-3$ & \low  & 4.3(7)     &  0. ~(fixed) & 17.0(8)  &  122 / 114  \\
    ~       &  LF   & 0.071(2)   &  6.4(26)     &  2.8(5)  &  ~  \\
    ~       & \LFT  & 0.131(3)   &  5.5(1.9)    &  3.4(4)  &  ~  \\
    \tableline
    ~       & break & 0.026(3)  &  0.32(9)     &  18.7(2.0) & ~ \\
    $3-10$  & hump  & 0.144(11) &  0.27(13)    &  25.9(2.0) & 119 / 112 \\
    ~       & \low  & 3.3(6)    &  0. ~(fixed) &  31.0(1.6) & ~ \\
    \cutinhead{NuSTAR-1}
    ~      & break & 0.024(4)  &  0.22(11)    &  19. (2)   & ~ \\
    $3-10$ & hump  & 0.19(5)   &  0. ~(fixed) &  25.2(1.9) & 94 / 92 \\
    ~      & LF    & 0.062(3)  &  3.6(1.7)    &  6.7(1.1)  & ~ \\
    ~      & \LFT  & 0.123(4)  &  6. (3)      &  6.0(1.2)  & ~ \\
    \tableline
    $10-30$ & break & 0.041(6) & 0. ~(fixed)  & 28.2(1.3) & 86 / 77 \\
    \cutinhead{NuSTAR-2}
    ~      & break  &  0.05(2)   &  0.10(20)   &  16. (4) & ~ \\
    $3-10$ & hump   &  0.32(14)  &  0. ~(fixed) &  15. (5) & 43 / 55 \\
    \tableline
    $10-30$ & break & 0.054(13)  &  0. ~(fixed) &  15.2(1.1) & 59/56 \\
    \tableline
	\end{tabular} \\
\end{table*}

The $0.4-3$ keV power spectrum could be fitted with 5 Lorentzian
components, three of which are broad noise terms and two are QPOs
(see Table \ref{tab:fit} for details). The centroid frequencies of
the QPO differ by a factor of two within the measurement uncertainty,
suggesting they are harmonically related.

While there is an indication the two QPOs are present in
the harder $3-10$ keV energy band as well, the power spectrum could be
adequately described using just three noise terms.  These noise
components are equivalent to the ones seen in the lower energy band,
but have systematically larger amplitudes.  Including the two QPOs
with frequencies fixed at their previous position, however, does give
a marginally significant detection. 
To determine if the QPOs are not independently resolved because their 
amplitudes drops relative to the band-limited noise terms, or because of a decrease in
signal-to-noise I investigate the dependence of the power spectrum on
energy in more detail. 

\begin{figure}[t]
    \centering
    \includegraphics[width=\linewidth]{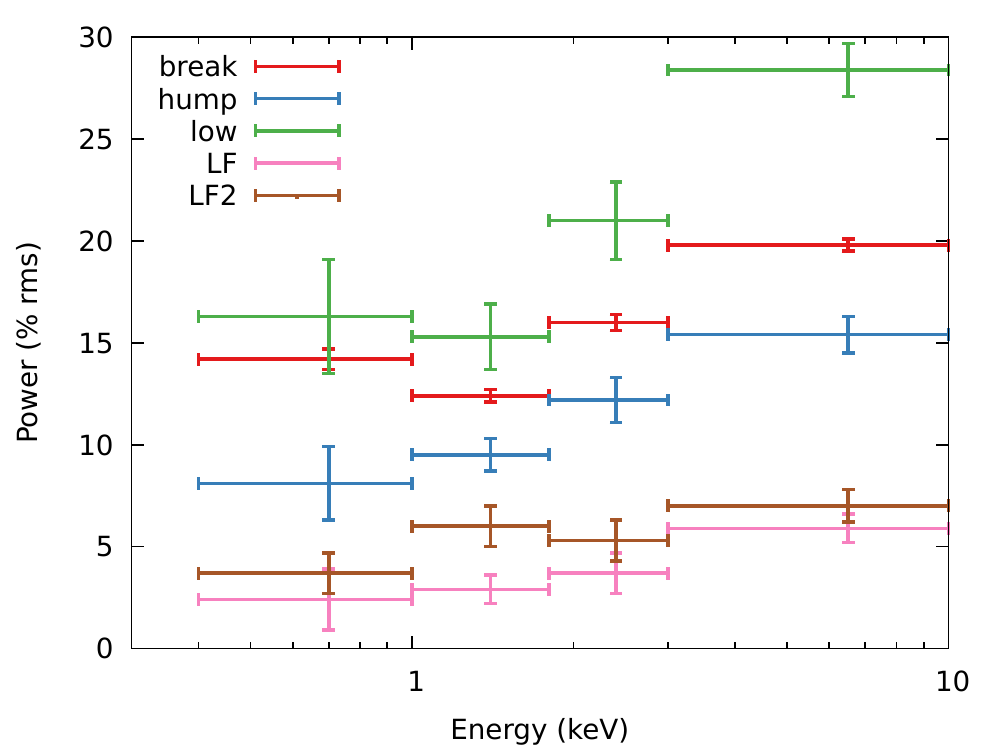}
    \caption{%
        Lorenztian component rms amplitudes as a function of energy.
    }
    \label{fig:amp.spec}
\end{figure}

Power spectra are constructed for 4 bands in the $0.4-10$ keV energy
range. Each power spectrum is fit with the 5 component
multi-Lorentzian model described above. To measure the energy
dependence the amplitudes are left to vary while all other
parameters are kept fixed.  As shown in Figure \ref{fig:amp.spec} all five
power spectral components show increasing amplitude as a function of
energy. This indicates that the lower signal-to-noise in the $3-10$ keV 
band is due to the lower count rate, rather than a decrease is signal
amplitude. 

The second \xmm observation (XMM-2) shows a similar power spectral
shape as seen in the first observation, but with all features
shifted to a slightly lower frequency  and higher rms amplitude 
(Figure \ref{fig:pds12}, second panel). The integrated power between
$0.01$ and $10$ Hz in the $0.4-3$ keV band stands at $27\%$ rms.
This power spectrum could again be fitted with a 5 component
Lorentzian model, the details of which are shown in Table
\ref{tab:fit}. The $3-10$ keV band shows a total integrated rms
amplitude of $44\%$ rms, and could be adequately described
with just the three broad noise terms.

For both XMM observations I also construct power spectra for light
curve segment lengths of $\sim1000$ seconds, so that the frequency range
extends down to $\sim1\E{-3}$ Hz. However, neither observation shows
additional power spectral features at very low frequencies.

For the first \nustar observation (NuSTAR-1) I first consider the $3-10$ keV energy band
that is also covered by \xmm. 
There is significant power between $0.01$ and $1$ Hz, with an 
integrated power of $\sim40\%$ rms. There may be an additional
feature $1$ Hz, however, the uncertainties there become to large 
to constrain the power density. 
The spectrum can be adequately fitted with two broad Lorentzian 
profiles and two QPOs (see Figure \ref{fig:pds12}, third panel). 
The band-limited noise in the \nustar power spectrum
is consistent with the two lower frequency noise terms seen in the
XMM-2 observation. 
The two QPOs seen in the \nustar data again differ 
in frequency by a factor of two within their uncertainties. In terms
of coherence and amplitude they are comparable to the QPOs seen in the \xmm
data, whereas their frequencies have slightly lower values.

In addition to the $3-10$ keV band I also construct power spectra
for the $10-30$ keV energy band. In this higher energy band the
power spectrum can be described using a single broad noise term, while
no QPOs could be resolved. 

    The second \nustar observation (NuSTAR-2) has a higher count rate, but a much
    shorter exposure (30 vs 60 ks), such that the signal-to-noise ratio
    of this data is lower. The $3-10$ keV power spectrum can be adequately
    described with two broad noise terms (Figure \ref{fig:pds12}, bottom
    panel). There is an indication of a narrow feature at $0.3$ Hz, but 
    with a significance of $\leq2\sigma$ it does not qualify as a
    detection.

\section{Discussion}
\label{sec:discussion}
    I have analyzed the X-ray variability of the accreting millisecond
    pulsar \maxi and found that the low frequency ($<10$ Hz) part of
    its power spectra can be described with two or three band-limited
    noise terms and two narrow QPOs.

\subsection{Feature identification} 
    The overall morphology of the power spectrum is reminiscent of an
    extreme island state atoll source \citep{Klis2006}.  For such a
    state, the three noise terms may be identified
    as the \textit{break}, \textit{hump}, and \textit{\low}
    components, each having a progressively higher frequency.
    The two QPOs, seen atop the hump component, may then
    be identified as the Low-Frequency (LF) QPO and its harmonic
    (LF$_2$, \citealt{Straaten2003, Altamirano2005}).
    The energy spectrum of \maxi is relatively hard \citep{Sanna2017a},
    which is consistent with an atoll source in this source state.
    The power spectral components, as shown in Table \ref{tab:fit},
    have therefore been labeled according to this terminology. 

    A notable difference between \maxi and the wider class of atoll
    sources is that the frequencies measured in this work are about
    an order of magnitude lower than what is usually observed (see
    e.g. \citealt{Doesburgh2017}).  Meanwhile, the observed fractional
    variability is remarkably high.  

    While extreme, these properties are in line with the source state evolution
    seen in atoll sources. As the energy spectrum increases in hardness, the
    power spectral features are observed to shift to lower frequencies, while
    broadening and increasing in amplitude \citep{Klis2006}.  In this context
    the source state of \maxi would represent an outlier along the evolutionary
    track, suggesting that this system is accreting in a geometry not normally
    accessible by regular atoll type objects.

\begin{figure}[t]
        \centering
        \includegraphics[width=\linewidth]{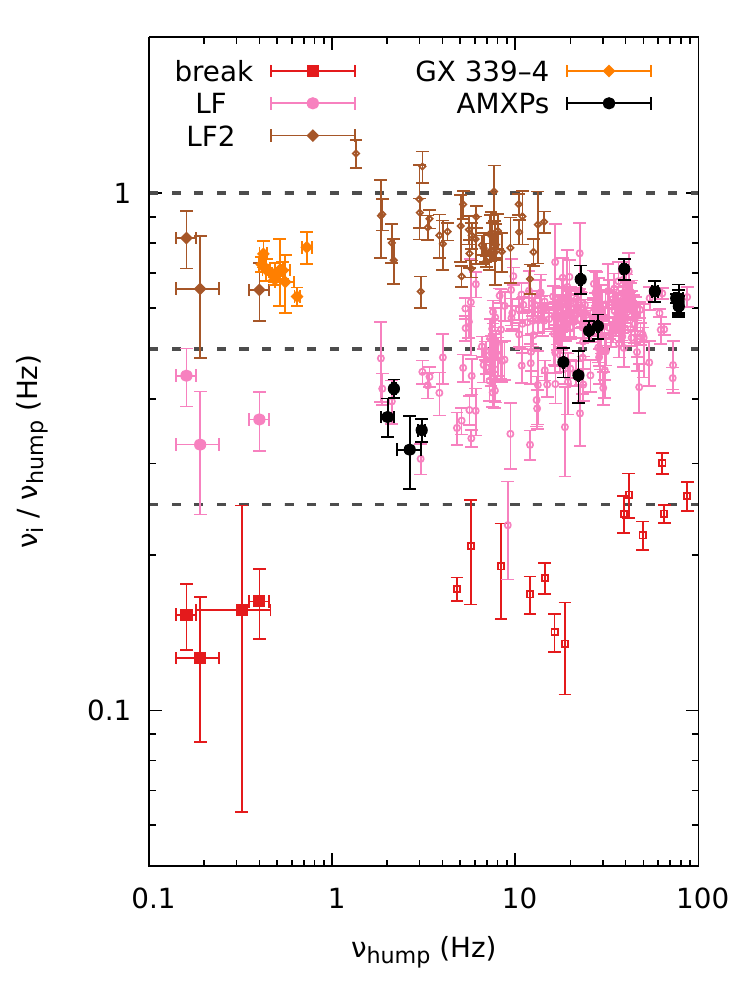}
        \caption{%
            Frequency ratios $\nu_i/\nu_{\rm hump}$ as a function
            of the hump frequency, for \maxi (large solid symbols to the left)
            as compared to a sample of atoll sources (small open symbols).
            Orange symbols give the \LFT frequencies measured in the
            black hole binary GX $339-4$, and black symbols give LF
            frequencies measured in other AMXPs.
        }
        \label{fig:ratios}
    \end{figure}

    The identification of power spectral features in neutron star
    sources is usually confirmed by considering the frequency
    relations as a function of the highest frequency component in
    the spectrum. However, with the high frequency end of the power 
    spectrum being poorly constrained, this is approach is not 
    possible for \maxi. 
    Instead I test the frequency relation of the observed features
    with respect to the hump component. In atoll sources these
    features have nearly fixed frequency relations
    \citep{Straaten2002, Altamirano2008b}, so that their frequency
    ratios are approximately constant \citep{Doesburgh2017}.  Hence
    the frequency ratios of the measured components relative to the
    hump component frequency are shown in Figure \ref{fig:ratios}.
    While the absolute frequencies in \maxi are low, the frequency
    ratios are consistent with those in other atoll sources, and
    with those of other AMXPs as well. This strongly supports 
    the identification of these features as LF QPOs.

    An alternative interpretation of the QPOs may be that they are
    related the mHz QPOs \citep{Revnivtsev2001}, which are attributed
    to marginally stable burning of accreted material on the neutron
    star surface \citep{Heger2007}. While the luminosity of \maxi is 
    in the regime where such QPOs are observed, there are several
    issues with this interpretations. Both the QPO frequencies and
    amplitudes in \maxi are higher by a factor of a few than is usual 
    for mHz QPOs \citep{Altamirano2008c,Linares2012}, and so far no 
    type I X-ray burst has been observed in \maxi. More importantly, the 
    QPO amplitudes show a steep increase as a function of energy, 
    strongly disfavoring a scenario in which they originate from 
    the neutron star surface.

    Yet another possibility might be that periodic obscuration along
    the line of sight is giving rise to these QPOs. However, again,
    this seems unlikely. Dipping QPOs tend to have higher frequencies
    and amplitudes than what is observed in \maxi \citep{Homan1999}. 
    What's more, the rms amplitude of a dipping QPO is expected to
    be independent of energy, which does not correspond with the
    observations. 

\subsection{IGR J00291+5934}
    There is only one other neutron star system that shows a
    power spectrum that is
    similar to the one observed in \maxi. That source 
    is the 599 Hz accreting millisecond pulsar IGR J00291+5934
    (IGR J00291; \citealt{Linares2007}).

    Observed with \textit{RXTE} in the $2-20$ keV band, the power
    spectrum of IGR J00291 shows a $0.01-100$ Hz integrated power of
    $40-60\%$ rms \citep{Linares2007}. The break, hump, 
    and low components all have comparable frequencies and amplitudes 
    to those seen in \maxi.
    Additionally, two very low frequency QPOs have been detected
    in IGR J00291. With
    frequencies of $22$ and $44$ mHz, however, they sit atop
    the break component rather than the hump component, and hence
    cannot be identified as LF QPOs such as those in \maxi.

    A very prominent 8 mHz flaring has also been detected in IGR
    J00291 \citep{Ferrigno2017}. This feature is not seen \maxi,
    but if it is due to `heartbeat' variability
    \citep{Belloni2000,Altamirano2011}, as suggested
    by \citet{Ferrigno2017}, then such flaring may be a transient 
    phenomenon and \maxi may in fact be a good candidate to monitor
    for such variability.

    The similarity of their broadband power spectra can be taking
    to indicate that both \maxi and IGR J00291 are accreting in a comparable 
    source state. That source state is then characterized by an highly
        energetic Comptonizing medium and a large fractional variability. As noted by
        \citet{Linares2007}, these properties are naturally explained if the
        optically thick accretion disk is truncated at a relatively large
        radius. An extended inner flow would then set the high fractional
        amplitude and the energy dependence , while the comparatively cool
        outer disk sets the low dynamical frequency \citep{Churazov2001,
        Gilfanov2003}. How such a configuration would interact with the stellar
        magnetic field remains unclear (see \citealt{Patruno2016a} or a
        discussion). 
    
    There is no obvious observable parameter indicating why these particular
    sources would have a larger truncation radius, or more generally such
    an extreme source state. 
    Both systems are millisecond pulsars, but have somewhat different
    properties \citep{Sanna2017a, Patruno2016b}. Their spin frequencies
    differ by nearly a factor of two. With an orbital period of $2660$
    s \maxi is in a tighter orbit than IGR J00291, which has a period 
    of $8844$ s. The mass limits on their companions is similar, and 
    both have accretion have luminosities of $\sim10^{36}$ erg s$^{-1}$ 
    \citep{Falanga2005,Sanna2017a}. However, the same can be said
    for other accreting millisecond pulsars. For instance, SAX
    J1808.4--3658 is more similar to IGR J00291 than \maxi, and yet
    the power spectra of that SAX J1808.4--3658 are far more consistent 
    with atoll sources at those luminosities \citep{Bult2015b}.

    Another peculiar property of \maxi is the duration of its outburst.
    Unlike IGR J00291, which shows a few week long outbursts every
    four to five years, \maxi has been continuously accreting for at least
    one year. From the two \nustar observations it follows that during
    this time the source state has remained unchanged. The only other
    AMXP to have shown such a long outburst is the 377 Hz intermittent pulsar
    HETE J1900.1--2455 \citep{Kaaret2006}, which remained active for about 
    ten years. The power spectrum of that source, however, is that of a regular (extreme) 
    island state atoll source \citep{Papitto2013}. This would suggest
    that neither intermittency nor outburst duty cycle has much bearing
    on the stochastic variability of these pulsars. 

    Potentially, the accretion geometry of \maxi and IGR J00291 can be
    accounted for by the detailed properties of the neutron star, 
    depending on magnetic field strength, stellar mass, as well as 
    the spin and magnetic alignment angles. While such properties
    are difficult to measure, detailed pulse profile modeling for 
    IGR J00291 and \maxi may provide further clues. 

\subsection{The LF QPOs}    
    The LF QPOs observed in \maxi have the lowest frequency seen
    in any accreting neutron star system. Instead their frequency
    range is more similar to those of the Low-Frequency QPOs observed 
    in low-hard state black hole binaries  \citep{KLeinWolt2008}. This
    is illustrated in Figure \ref{fig:ratios}, which also shows
    frequency ratios for the black hole binary GX $339-4$.

    For black hole binaries there is strong evidence that LF QPOs (or
    type C QPOs in black hole terminology) are caused by
    Lense-Thirring precession \citep{Ingram2016,Ingram2017}. In this
    model the accretion disk consists of two components; an inner, hot
    accretion flow emitting a hard Comptonized spectrum, and an outer
    thin disk emitting a softer thermal spectrum. If the black hole
    spin is misaligned with the accretion disk, frame dragging effects
    apply a torque on the hot flow, causing it to precess as a solid
    body \citep{Ingram2009}. This precession is argued to cause the LF
    QPOs.
    Qualitatively, this model is in good agreement with the power
    spectral features of \maxi and the hard energy spectrum of their
    amplitudes. 

    If neutron star systems have a precessing inner accretion flow 
    like those predicted for black holes, then their LF QPO 
    frequencies are not straightforwardly related to the Lense-Thirring 
    precession frequency \citep{Altamirano2012, Bult2015b, Doesburgh2017}.
    For instance, in neutron star systems the stellar oblateness
    introduces an additional torque on the disk, giving rise to a
    retrograde precession term \citep{Morsink1999}.  For accreting
    millisecond pulsars this situation is complicated further by the
    stellar magnetosphere, which truncates the accretion disk and has
    been proposed to introduces a third, magnetic torque
    \citep{Lai1999, Shirakawa2002}, leading to a second retrograde
    precession term. 

    Determining how, exactly, these various torques set the LF QPO 
    frequency has the potential of yielding constraints on how the
    magnetosphere interacts with the disk, and of the neutron star
    properties itself. 
    The black-hole like power spectrum of \maxi make this source
    an ideal target for such a study, for instance with
    the \textit{Neutron Star Interior Composition
    Explorer} (NICER, \citealt{Gendreau2012}), which is scheduled for
    launch in 2017. 

\acknowledgments
I would like to thank Marieke van Doesburgh for sharing some of the data 
behind Figure 4, and the referee for constructive feedback that helped improve
the paper. This work was was supported by an NPP fellowship at NASA Goddard 
Space Flight Center.

\appendix
\input{cospec}

\bibliographystyle{aasjournal}

\end{document}

%% file: cospec.tex
\section{Co-spectrum statistics}
\label{app:nustar}
    When a detector has more than one focal plane module, a single
    observation produces several concurrent time series. These time
    series may the be combined by either adding them, or by correlating
    them. In the following the statistical properties of each approach
    are derived and then compared. 

\subsection{Coherent averaging}
    Consider a discrete time series $x_j$ containing a total of $N_\gamma$
    counts. It is well known that for a time series containing only Poisson 
    noise, the elements of the Fourier transform, $X_k$, will be distributed 
    as a complex normal variates with zero mean and variance $N_\gamma/2$ 
    \citep{Leahy1983b}. If I let
    \begin{equation}
    \label{app:signal}
        Z_k = \sqrt{\frac{2}{N_\gamma}} X_k = A_k e^{i \phi_k} + Z_{\rm noise},
    \end{equation}
    be a normalized Fourier transform of a signal in the presence of Poisson noise,
    then $A_k^2$ gives the signal Leahy power, $\phi_k$ is its underlying phase,
    while $Z_{\rm noise}$ is a complex-valued noise term with independent standard 
    normal variates in both its real and imaginary parts.
    
    Given an ensemble of $M$ concurrent measurements, the Fourier transforms 
    can be averaged coherently (dropping the frequency index $k$ for convenience)
    \begin{equation}
        \avg{Z}_M = A e^{i \phi} + \avg{Z_{\rm noise}}_M,
    \end{equation}
    so that the variance reduces to $1/M$. By definition the sum of 
    squared standard normal variates follows a chi-squared 
    distribution. Similarly, the sum of squared normal variates that 
    have a non-zero mean, $\mu_n$, and unit variance will 
    follow a \textit{non-central chi-squared} distribution, $\overline \chi^2(\nu,\lambda)$, 
    characterized by $\nu$ degrees of freedom and non-centrality
    \begin{equation}
        \lambda = \sum_{n=1}^{\nu} \mu_n^2.
    \end{equation}
    It then follows that the power of $\avg{Z}_M$ is distributed as 
    $\overline\chi^2(2,MA^2)$ scaled by a factor of $1/M$.
    
    Given an ensemble of $L$ coherently averaged powers, the combined averaged 
    power is distributed as
    \begin{equation}
    \label{app:coh.distri}
        f_P(y) = \frac{1}{LM}\overline\chi^2\rbr{ \frac{y}{LM} \mid \nu=2L , \lambda=LMA^2 },
    \end{equation}
    which has a mean of
    \begin{equation}
        \mu_P = \frac{2}{M} + A^2,
    \end{equation}
    and variance
    \begin{equation}
    \label{app:coh}
        \sigma^2_P = \frac{4+4MA^2}{LM^2}.
    \end{equation}
    Note that for a signal that is only averaged incoherently ($M=1$),
    eq. \ref{app:coh.distri} reduces to the result derived by
    \citet{Groth1975} for a signal in the presence of noise.
    For a pure noise signal ($A = 0$) these results reduce further
    to the familiar chi-squared distribution with $2L$ degrees of freedom, and 
    yields a Poisson noise level of $2$ \citep{Klis1989}.

\subsection{Correlating concurrent series}
    For an ensemble of $M$ concurrent time series (as eq.
    \ref{app:signal}) there are $\tilde M = \binom{M}{2} =
    \rbr{M-1}M/2$ unique pairs, each of which can be correlated to
    give an estimate of the co-spectrum. For such a pair, say $x_a[j]$
    and $x_b[j]$, the co-spectrum at a particular frequency index is 
    given as
    \begin{equation}
        C_{ab} = \mbox{Re}\br{ Z_a Z_b^* } = R_a R_b + I_a I_b,
    \end{equation}
    where $R_a$ and $R_b$ are normal variates with mean ${A\cos\phi}$
    and unit variance. The terms $I_a$ and $I_b$ are similarly
    distributed with a mean of ${A\sin\phi}$.
    It is useful to define the product $Y_{ab} = R_aR_b$, for which
    the expectation and variance are given as
    \begin{align}
        \Ex{Y_{ab}}  &= \Ex{R_a}\Ex{R_b} = A^2 \cos^2\phi, \nonumber \\
        \Var{Y_{ab}} &= \Ex{R_a^2}\Ex{R_b^2} -
                        \rbr{\Ex{R_a}\Ex{R_b}}^2, \nonumber \\ 
                    &=1 + 2 A^2 \cos^2\phi. \label{app:prod.var}
    \end{align}
    When averaging the product $Y$ over all $\tilde M$ correlation pairs
    the mean will reduce to this expectation. For the variance is
    important to realize that the pairs are not all independent.
    Specifically, the two products $Y_{ab}$ and $Y_{ac}$ have a
    covariance of
    \begin{align}
        \mbox{Cov}\br{Y_{ab}, Y_{ac}} 
        &= \Ex{R_a^2 R_b R_b } - \Ex{Y_{ab}}\Ex{Y_{ac}}, \nonumber \\
        &= A^2 \cos^2 \phi. \label{app:covar.y}
    \end{align}
    If I let $\Gamma_{nm}$ be the full covariance matrix of the $\tilde M$
    correlation pairs, then the variance of the averaged product term will be
    \begin{align*}
        \Var{\avg{Y}_{\tilde M}} 
          &= \frac{1}{\tilde M^2} 
            \sum_{m=0}^{\tilde M-1} 
                \sum_{n=0}^{\tilde M-1} 
                    \Gamma_{nm} \\ 
          &= \frac{1}{\tilde M^2} \sbr{
                \sum_{m=0}^{\tilde M-1} \Gamma_{mm}
        + \sum_{m=0}^{\tilde M-1} \sum_{\substack{n=1 \\ n \neq m}}^{\tilde M-1} 
        \Gamma_{nm} }
    \end{align*}
    Each term in the first summation is given by eq
    \ref{app:prod.var}.
    Because each of the $\tilde M$ pairs is correlated with 
    $2(M-2)$ other pairs, the number of non-zero terms in the second 
    sum is $M(M-1)(M-2)$, with each contribution given by eq.
    \ref{app:covar.y}. The variance then reduces to
    \begin{align*}
        \Var{\avg{Y}_{\tilde M}} = 
        \frac{1 + 2 (M-1) A^2 \cos^2\phi}{\tilde M}
    \end{align*}

    The product $I_a I_b$ has similar averaged properties, but with
    the cosine terms replaced by sines. Since all sine and cosine
    terms are squared, it is easy to see how they drop out when
    the two averaged products are added to a single average
    co-spectrum. 
    
    Finally, considering an ensemble of $L$ separate observations,
    the co-spectrum estimator gives a mean and variance of
    \begin{align}
        \mu_C &= A^2, \\
        \sigma_C^2 &= \frac{4 + 4(M-1) A^2}{ L M (M-1) }  \label{app:cor}
    \end{align}

\subsection{Comparing convergence}
    Given a sample of $K$ observations with a two detector system I can
    distinguish three averaging scenarios.  First, averaging in the time domain
    gives a standard deviation from eq. \ref{app:coh} using $M=2$ and $L=K$.
    Second, averaging by correlating in the Fourier domain gives a standard
    deviation from eq. \ref{app:cor} using $M=2$ and $L=K$.  Third, averaging
    powers gives a standard deviation as in eq. \ref{app:coh} with $M=1$ and
    $L=2K$.  For pure noise the ratios of these uncertainties relate as
    $1:\sqrt{2}:\sqrt{2}$, indicating that time domain averaging gives the best
    detection sensitivity, whereas correlating the concurrent time series gives the same
    sensitivity as combining their power spectra.